\documentclass[a4paper]{jpconf}
\usepackage{graphicx}
\usepackage{amssymb} 
\usepackage{amsthm} 
\usepackage{lineno}
\usepackage{ulem}
\usepackage{color}
\usepackage[usenames,dvipsnames,svgnames,table]{xcolor}
\begin{document}
\title{QCD bulk thermodynamics and conserved charge fluctuations with HISQ fermions}
\author{Christian Schmidt (for HotQCD\footnote{Collaborations members are: A. Bazavov, T. Bhattacharya, M. Buchoff, M. Cheng, N. Christ, C. DeTar, H.-T. Ding, S.Gottlieb, R. Gupta, P. Hegde, U. Heller, C. Jung, F. Karsch, E. Laermann, L. Levkova, Z. Lin, R.~Mawhinney, S. Mukherjee, P. Petreczky, D. Renfrew, C. Schmidt, C. Schroeder, W. S\"oldner, R. Soltz, R.~Sugar, D. Toussaint, P. Vranas.} and BNL-Bielefeld\footnote{Collaboration members are: A. Bazavov, H.-T. Ding, P. Hegde, O. Kaczmarek, F. Karsch, E. Laermann, 
S.~Mukherjee, P. Petreczky, C. Schmidt, D. Smith, W. Soeldner,  M. Wagner.} Collaborations)}
\address{Fakult\"at f\"ur Physik, Universit\"at Bielefeld, Postfach 100131, D-33501 Bielefeld, Germany}
\ead{schmidt@physik.uni-bielefeld.de}
\begin{abstract}
After briefly reviewing recent progress by the HotQCD collaboration in studying the 2+1 flavor QCD equation of state, we will focus on results on fluctuations of conserved charges by the BNL-Bielefeld and HotQCD collaborations. Higher order cumulants of the net-charge distributions are increasingly dominated by a universal scaling behavior, which arises due to a critical point of QCD in the chiral limit. Considering cumulants up to the $6^{th}$ order, we observe that they generically behave as expected from universal scaling laws, which is quite different from cumulants calculated within the hadron resonance gas model. Taking ratios of these cumulants, we obtain volume independent results that can be compared to the experimental measurements. We will argue that the freeze-out chemical potentials and the freeze-out temperature, usually obtained by a HRG model fit to the measured hadronic yields, can also be obtained in a model independent way from ab-initio lattice QCD calculations by utilizing observables related to conserved charge fluctuations. Further, we will show that the freeze-out strangeness and electric charge chemical potentials can be fixed by imposing strangeness neutrality and isospin asymmetry constraints in the lattice QCD calculations, in order to accommodate conditions met in heavy ion collisions.
All results have been obtained with the highly improved staggered quark action (HISQ) and almost physical quark masses on lattices with temporal extent of $N_\tau=6,8,10,12$.
\end{abstract}

\section{Introduction}
\begin{figure}[htbp]
\begin{center}
\begin{minipage}{0.4\textwidth}
\begin{center}
\includegraphics[width=0.99\textwidth]{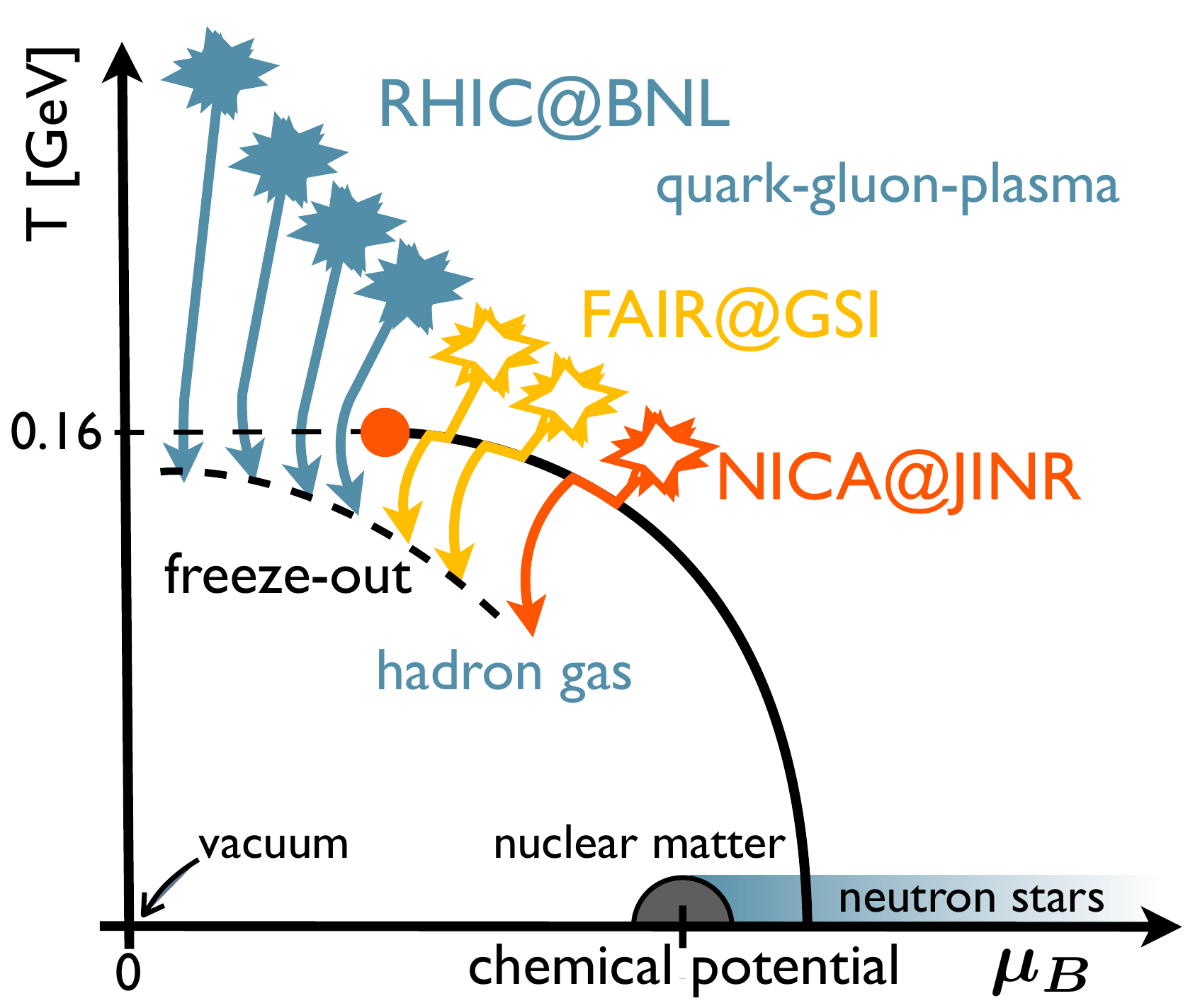}
\end{center}
\end{minipage}
\end{center}
\caption{Generic phase diagram of QCD, based on model calculations and model independent symmetry arguments.  Also indicated are the approximate initial conditions of the fireballs that are created in heavy ion experiments as well as their hydrodynamic trajectories on which they evolve, until they reach chemical freeze-out. \label{fig:pdiag}}
\end{figure}

Calculating the phase diagram of strongly interacting matter is one of the most important and outstanding problems of non-perturbative QCD. A generic plot of the QCD phase diagram based on model calculations and model independent symmetry arguments is shown in Fig.\ref{fig:pdiag}.
To clarify whether or not there is a critical point in the phase diagram is relevant for cosmology, astro physics and heavy ion collisions. Currently various experimental programs aim at probing the phase diagram at nonzero baryon number density, such as the beam energy scan at RHIC, NA61 at the SPS as well as future experiments at FAIR and NICA. In Fig.\ref{fig:pdiag} we have indicated the approximate initial conditions of the fireballs as they are created in some of those experiments. Also shown is a sketch of the trajectories on which the fireball evolves in the temperature ($T$) and baryon chemical potential ($\mu_B$) plane, during its expansion and cooling. It has been shown that this evolution can be very well described by hydrodynamics, with a surprisingly low viscosity \cite{Romatschke}. While the initial conditions have to be modeled \cite{Schenke}, the hydrodynamic evolution relies, at least up to first order, on the QCD equation of state (EoS) and two additional transport coefficients. The EoS can be obtained by first principle lattice calculations and we report on recents strategies of the HotQCD collaboration to calculate the EoS with the highly improved staggered quark (HISQ) action. Note that due to the notorious sign problem lattice QCD calculations are currently restricted to zero baryon chemical potential. However, the effect of a nonzero chemical potential can be taken into account by a controlled Taylor expansion in $\mu_B/T$ \cite{swansea,epj}.

We continue with a discussion of fluctuations of conserved charges such as net baryon number, electric charge and strangeness. As one expects a critical point in the phase diagram at nonzero density, fluctuation observables are the most natural choice to detect critical behavior in QCD. Their divergent behavior can give insight in the structure of the phase diagram and has been proposed as a signal for the critical point \cite{rajagopal}. Moreover, as we argue here, they can be used to extract the chemical freeze-out conditions as observed in heavy ion collisions \cite{PRL,Swagato}. After hadronization of the fireball, we expect a strongly interacting gas of hadrons with a fluctuating chemical decomposition. However, from a given point on the trajectory, the particle abundances stay fixed. This point is called chemical freeze-out. The coordinates of this point in the $(T,\mu_B)$-diagram are usually obtained by fitting the experimentally measured particle abundances to the hadron resonance gas (HRG) model. Those fits have been quite successful in the past \cite{HRG}. It has been shown that the freeze-out parameters that are obtained by HRG model fits from various experiments at different collision energies fall on a unique curve that can be parametrized as a function of the center of mass energy ($\sqrt{s_{NN}}$) \cite{fc}. 

\section{The highly improved staggered quark (HISQ) action}
\begin{figure}[htbp]
\begin{center}
\begin{minipage}{0.396\textwidth}
\begin{center}
\includegraphics[width=0.99\textwidth]{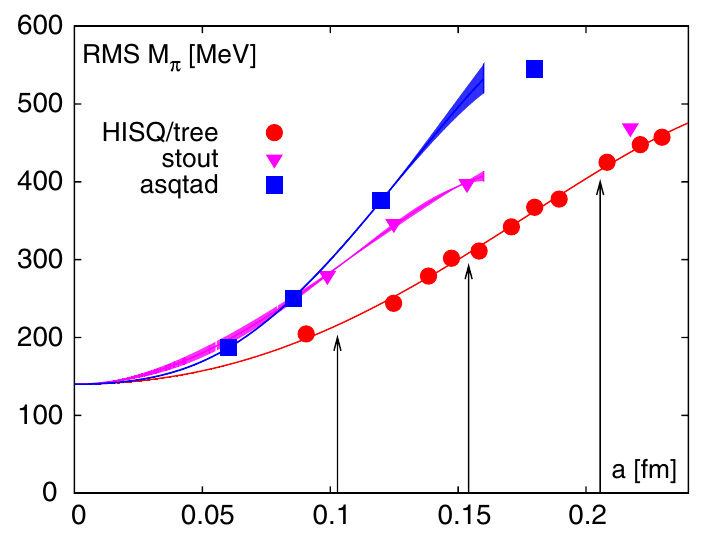}
\end{center}
\end{minipage}
\begin{minipage}{0.459\textwidth}
\begin{center}
\includegraphics[width=0.99\textwidth]{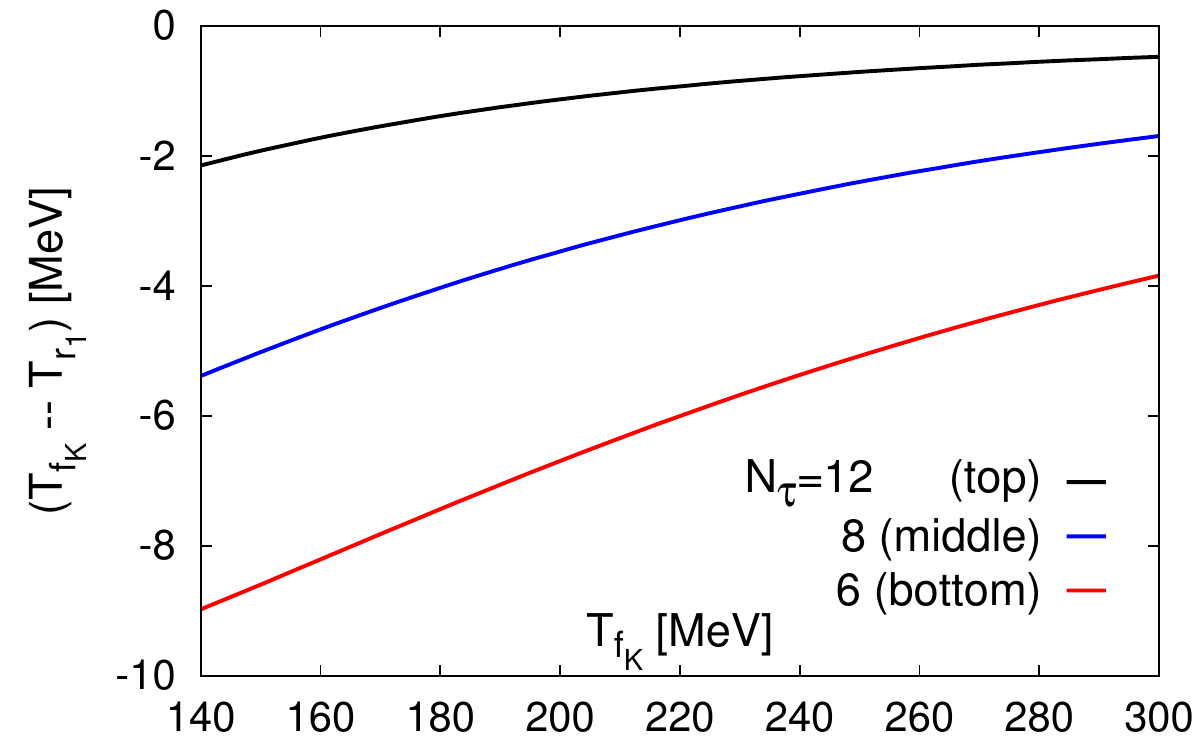}
\end{center}
\end{minipage}
\end{center}
\caption{Root-mean-square mass of the pion spectrum as function of the lattice spacing ($a$) for the HISQ/tree, stout and asqtad action (left). Arrows indicate the lattice spacings corresponding to our $N_\tau=12,8$ and 6 lattices (from left to right) at $T=200$ MeV. The right panel shows the temperature ambiguity as obtained from two different scales for $N_\tau=12,8$ and 6 lattices.  \label{fig:hisq}}
\end{figure}

Lattice QCD simulations utilize a discretized version of the QCD action, which at non-zero values of the lattice spacing is not uniquely defined. In fact, many lattice actions exist, which differ in the way they handle fermion doublers, to what extent the chiral symmetry is preserved and also in the amount of discretization effects they possess. A widely used class of lattice actions is given by various improved versions of so called staggered actions. They are numerically cheap and preserve a $U(1)$ subgroup of the $SU(N_f)_L\times SU(N_f)_R$ chiral symmetry that is manifest in the QCD Lagrangian at zero quark masses. Results that are presented here have been obtained with the HISQ action \cite{HISQ}, that strongly reduces lattice discretization effects over the naive staggered action in two perspectives. Firstly, the discretization effects stemming from the covariant derivative term are improved by incorporating in addition to the usual nearest neighbor discretization of the derivative a higher order three-link term (Naik-term). This term dramatically reduces discretization effects at high temperature. Secondly, the flavor (taste) symmetry breaking, which is inherent to staggered quark actions, is reduced by a two step smearing procedure, where the smeared links are projected back to the gauge group $SU(3)$ after the first step. The taste breaking results in an unphysical splitting of the pion spectrum at given lattice spacing, which then leads to an averaged pion mass (root-mean-square mass $m_\pi^{\rm RMS}$). Although the lightest pion mass (Goldstone pion) is kept fixed through out our simulations at different lattice spacings the $m_\pi^{\rm RMS}$ varies. It can be seen as a measure of the taste breaking effects and is shown in Fig.~\ref{fig:hisq} (left). 
As one can see, among the three compared staggered actions the HISQ action performs best. The improvement of the taste breaking is most important at low and intermediate temperatures and, as discused in Sec.~\ref{sec:HRG}, drastically modifies the electric charge fluctuations in the hadron gas phase. 

The results that are presented here have been obtained with (2+1)-flavor of HISQ fermions, with a strange quark mass ($m_s$) that was tuned to its physical value and light quark masses that are fixed by the quark mass ratio $m_l/m_s=1/20$. This is slightly heavier than the physical value $m_l/m_s\approx 1/27$. We calculate on lattices with temporal extent of $N_\tau=6,8,10,12$ and spacial extent $N_\sigma=4N_\tau$. For the $T=0$ calculations needed to renormalize the EoS calculations, we use $N_\tau\geq N_\sigma$.
The temperature scale was set by two independent experimentally accessible quantities, the Sommer scale \cite{Sommer} ($r_1$) and the Kaon decay constant ($f_K$). The ambiguity in the temperature obtained by those two quantities at finite $N_\tau$ is shown in Fig.~\ref{fig:hisq} (right).  

\section{Status of the EoS calculation}
\begin{figure}[htbp]
\begin{center}
\begin{minipage}{0.345\textwidth}
\begin{center}
\includegraphics[width=0.99\textwidth]{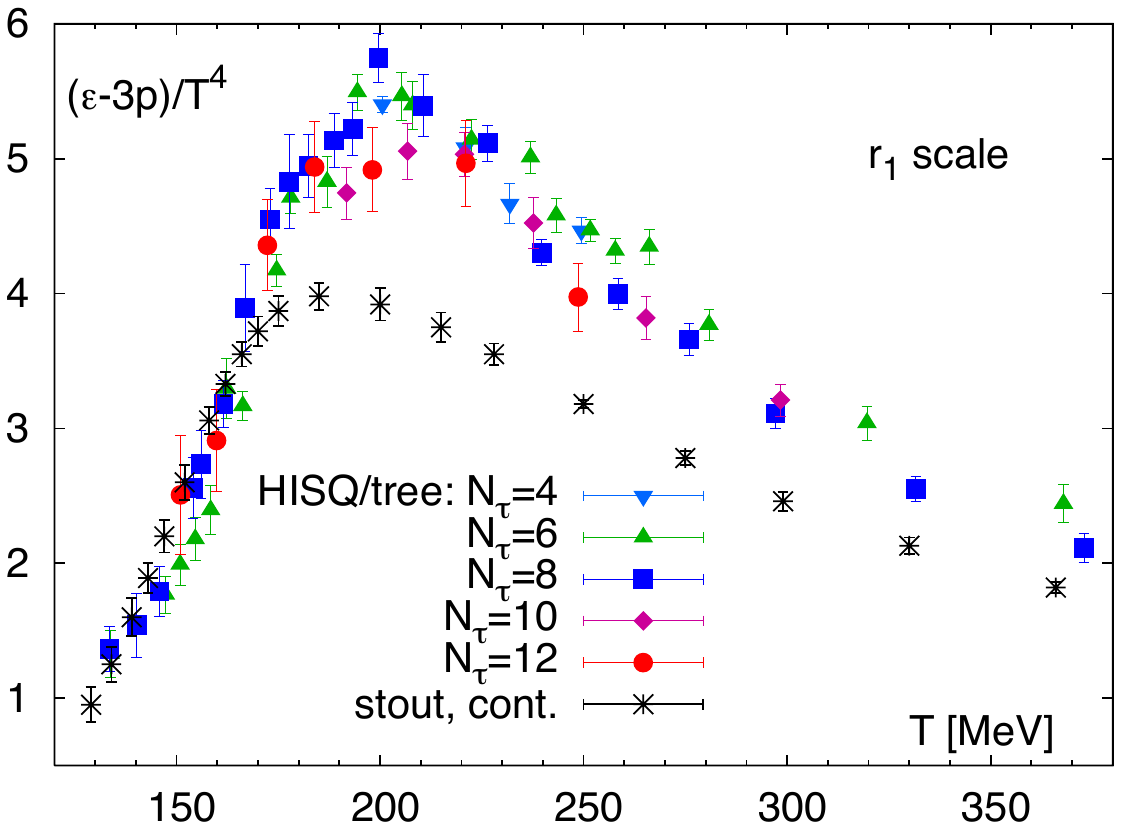}
\end{center}
\end{minipage}
\begin{minipage}{0.315\textwidth}
\begin{center}
\includegraphics[width=0.99\textwidth]{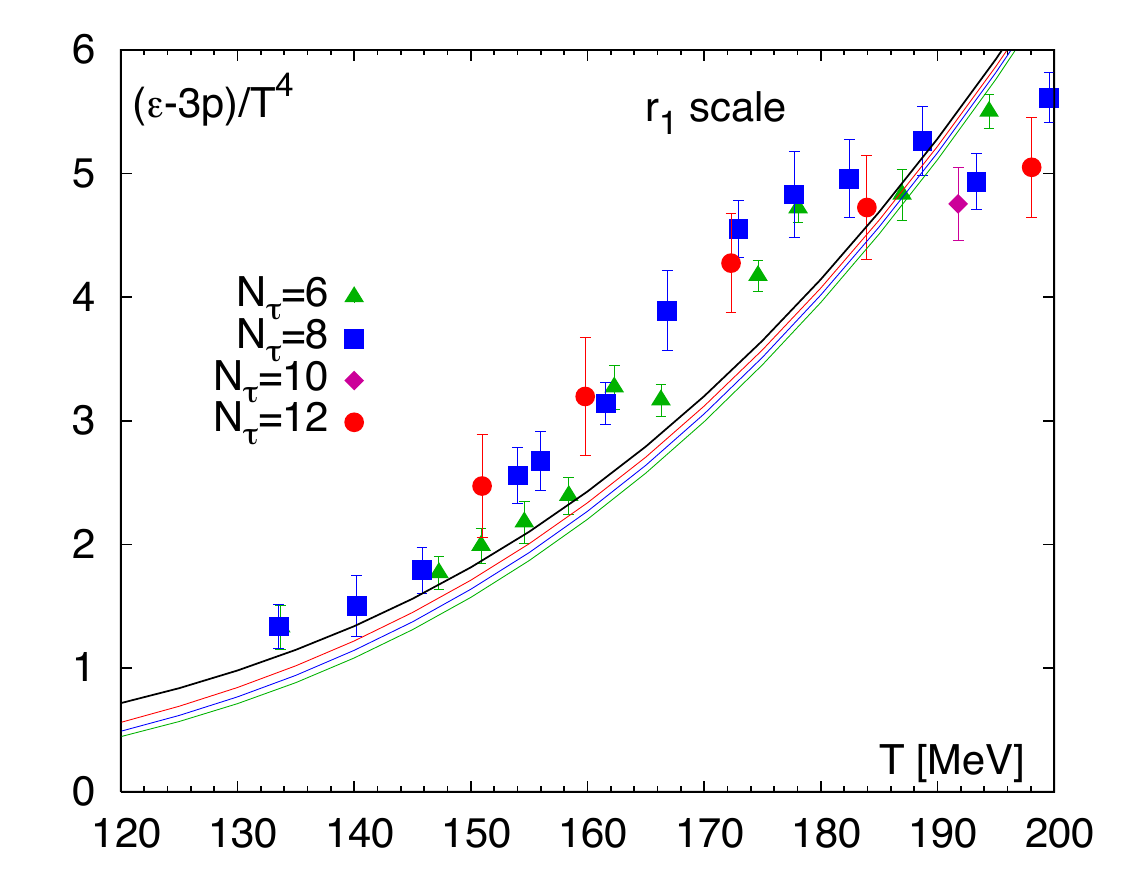}
\end{center}
\end{minipage}
\begin{minipage}{0.305\textwidth}
\begin{center}
\includegraphics[width=0.99\textwidth]{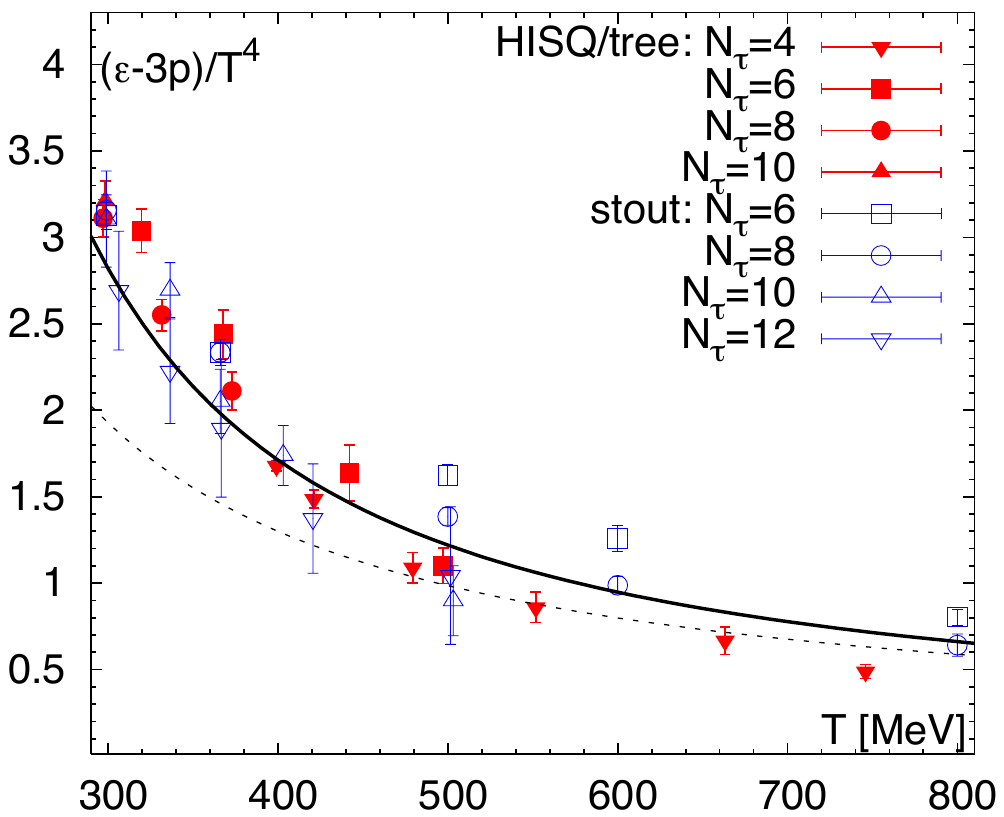}
\end{center}
\end{minipage}
\end{center}
\caption{Comparison of the HISQ interaction measure on $N_\tau=6,8,10$ and 12 lattices with the stout continuum estimate \cite{stout} (left). The HISQ interaction measure at low (middle) and high (right) temperatures. Solid lines in the middle panel indicate HRG results, dashed and solid lines in the right panel show 1-loop and 2-loop perturbative results, respectively. \label{fig:eos}}
\end{figure}

The quantity most convenient to calculate on the lattice is the trace anomaly in units of the fourth power of the temperature $\Theta^{\nu\nu}/T^4$. This is given by the derivative of $p/T^4$ with respect to the temperature, {\it i.e.}
\begin{equation}
\frac{\Theta^{\nu\nu}(T)}{T^4}\equiv\frac{\epsilon-3p}{T^4}=T\frac{\partial}{\partial T}(p/T^4).
\end{equation}
Since the pressure is given by the logarithm of the partition function, $p/T = V^{-1} \ln Z$, the calculation of the trace anomaly requires only the evaluation of straightforward expectation values. Note, however, that we have to subtract the corresponding $T=0$ result in order to remove the UV divergencies. {\it I.e.}, for each parameter set of couplings and masses we have to perform an additional $T=0$ calculation, which makes the EoS determination numerically very demanding.  All further thermodynamic quantities can be deduced from the trace anomaly, {\it e.g.} the pressure is obtained as
\begin{equation}
\frac{p(T)}{T^4}-\frac{p(T_0)}{T_0^4}=\int_{T_0}^{T}dT^\prime \; \frac{\Theta^{\nu\nu}(T)}{{T^\prime}^5}.
\label{eq:p}
\end{equation}
In Fig.~\ref{fig:eos} we show our current status of the $\Theta^{\nu\nu}(T)/T^4$ calculations from $N_\tau=6,8,10$ and 12 lattices \cite{Bazavov}. 
Also shown is a continuum estimate, obtained with a different staggered type action, the stout action, taken from Ref.~\cite{stout}. As one can see, results obtained with the two actions currently are not in agreement in the temperature region $180\lesssim T\lesssim 300$~MeV. The discrepancies are most pronounced at $T\approx 200$ MeV, where the trace anomaly has its maximum. Note, that the HISQ data is not yet extrapolated to the continuum, which may reduce or eliminate the discrepancy. It is, however, obvious that for the pressure that is obtained by the integral in Eq.~(\ref{eq:p}), we encounter differences of (10-15)\% for $T\gtrsim 300$~MeV between the present HISQ data at $N_\tau=12$ and the stout continuum estimate \cite{stout}.

The middle panel in Fig.~\ref{fig:eos} shows the low temperature region of the interaction measure. The solid lines indicate corresponding HRG results, where from top to bottom the pion mass was set to its physical value and to the root-mean-square mass $m_\pi^{\rm RMS}$ corresponding to our $N_\tau=12,10,8$ and 6 lattices. As one can see, the trace anomaly calculated in the HRG model is not very sensitive to the pion mass already at temperatures above 130 MeV. Furthermore, one sees that the lattice data approach the HRG model at $T\sim 150-160$ MeV.

Fig.~\ref{fig:eos} (right) shows the trace anomaly obtained with the HISQ and stout actions in the high temperature regime. Also shown here are the perturbative results at 1-loop and 2-loop order, indicated by the dashed and solid lines, respectively. We find good agreement with the lattice data already above $T\gtrsim 300$~MeV.   

\section{Conserved charge fluctuations}
We will now continue with the discussion of fluctuations of conserved charges, which can be measured in experiments and calculated on the lattice. They are well suited  to study critical behavior but can also be used to determine freeze-out conditions. 
\subsection{Cumulants of conserved charge fluctuations}
\begin{figure}[htbp]
\begin{center}
\begin{minipage}{0.293\textwidth}
\begin{center}
\includegraphics[width=0.998\textwidth]{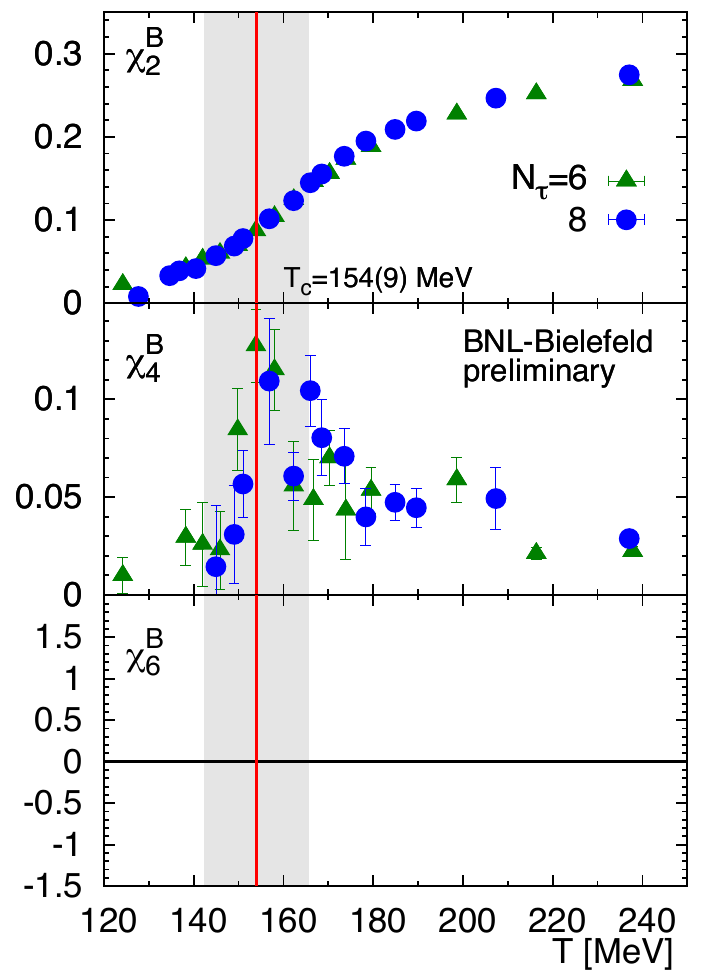}
\end{center}
\end{minipage}
\begin{minipage}{0.293\textwidth}
\begin{center}
\includegraphics[width=0.998\textwidth]{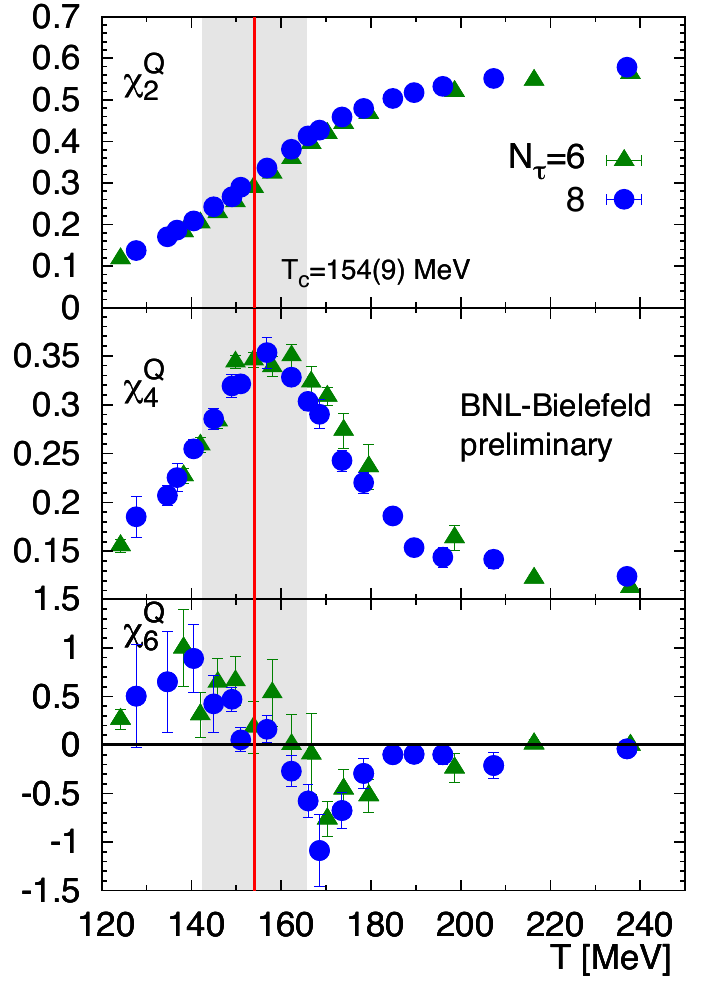}
\end{center}
\end{minipage}
\begin{minipage}{0.399\textwidth}
\begin{center}
\includegraphics[width=0.998\textwidth]{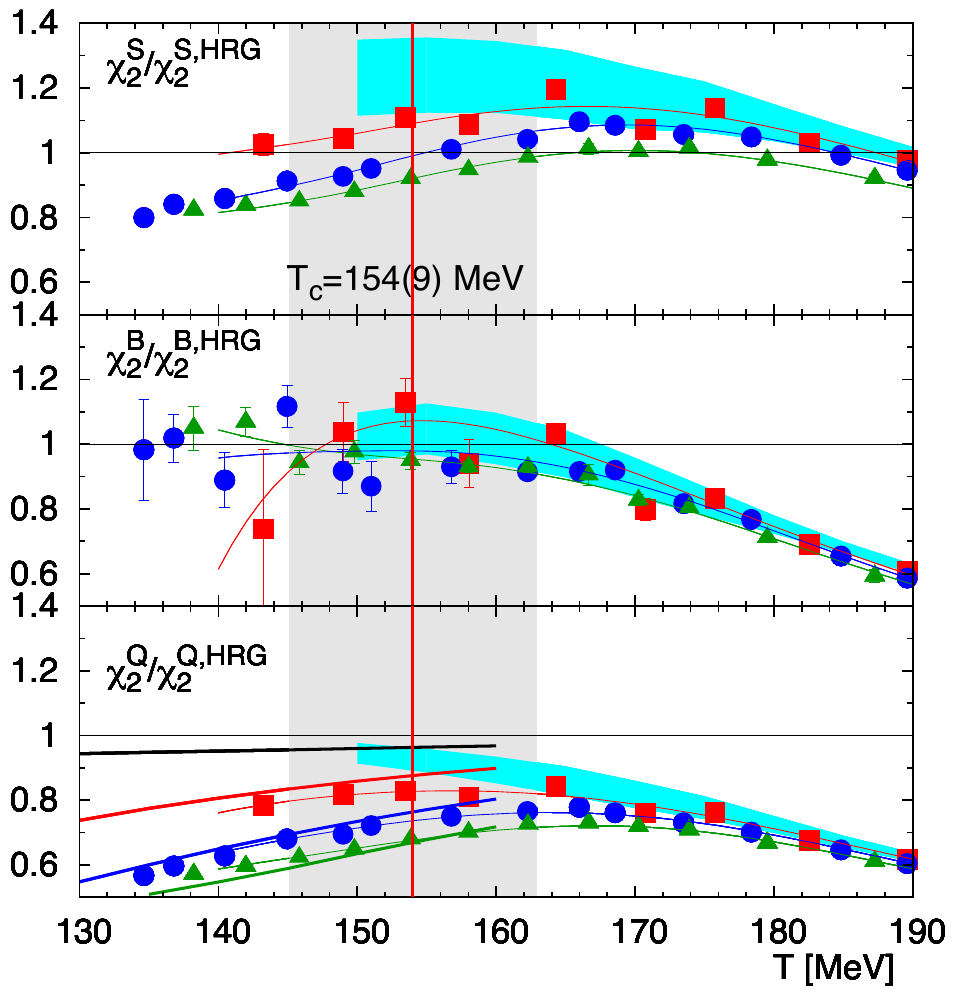}
\end{center}
\end{minipage}
\end{center}
\caption{Cumulants of net baryon number fluctuations (left) and net electric charge fluctuations (middle) up to the sixth order; quadratic cumulant of strangeness, baryon number and electric charge normalized by the HRG results (right). Different symbols denote different lattice spacings: $N_\tau=6$ (triangles), $8$ (circles) and $12$ (squares). Solid lines on the right panel indicate spline fits to the data. For the electric charge fluctuations additional solid lines indicate HRG results with a modified pion mass (see text for explanation). The vertical bar indicates the QCD transition temperature as obtained in \cite{Tc}. \label{fig:chi}}
\label{fig:harmony}
\end{figure}

On the lattice we calculate  derivatives of the partition function with respect to baryon ($B$), electric charge ($Q$) and 
strangeness ($S$) chemical potentials, which are also known as generalized susceptibilities and are defined as 
\begin{equation}
\left(VT^3\right)\cdot\chi^{BQS}_{ijk}(T)= \left(\partial^{i+j+k}\ln Z(T,\mu_B,\mu_Q,\mu_S)\right) \left/ \left(\partial \hat\mu_B^i \partial \hat\mu_Q^j \partial \hat\mu_S^k \right)\right.,
\end{equation} 
with $\hat\mu_X=\mu_X/T$ and $X=B,Q,S$. The lattice studies are performed at $\mu_X=0$ and sufficiently close to the thermodynamic limit. The generalized susceptibilities are intensive quantities. They can also be interpreted as Taylor expansion coefficients of $\ln Z$ and, furthermore, be related to the cumulants of the fluctuations of the net charges ($N_X$), which are measured in heavy ion collisions. {\it E.g.}, for the diagonal fluctuations one obtains
\begin{eqnarray}
\left(VT^3\right)\cdot\chi^X_2 &=& \left<\left(\delta N_X\right)^2\right>,\\
\left(VT^3\right)\cdot\chi^X_4 &=& \left<\left(\delta N_X\right)^4\right>-3\left<\left(\delta N_X\right)^2\right>^2,\\
\left(VT^3\right)\cdot\chi^X_6 &=& \left<\left(\delta N_X\right)^6\right>-15\left<\left(\delta N_X\right)^4\right>\left<\left(\delta N_X\right)^2\right>+30\left<\left(\delta N_X\right)^2\right>^3,
\end{eqnarray}
with $\delta N_X=N_X-\left<N_X\right>$. In Fig.~\ref{fig:chi} (left and middle) we show our current results for the 
diagonal fluctuations of net baryon number and electric charge, obtained with the HISQ action on lattices with temporal extent of $N_\tau=6$ and $8$. 

\subsection{O(4) critical behavior}
In the chiral limit of QCD with two massless quarks the transition from the hadronic world 
to the quark gluon plasma is presumably of second order (see Fig.~\ref{fig:pdiag}). 
At vanishing chemical potential the critical point is expected to be in the universality class of the three dimensional $O(4)$ symmetric spin model 
and is expected to persist even in the presence of a physical strange quark mass. It is illuminating to perform a universal scaling 
analysis that is connected with the critical point in the chiral limit. Here to leading order the chemical potential 
enters only in the temperature like scaling field
\begin{equation}
t=t_0^{-1}\left((T-T_0)/T_0 + \kappa_B \hat\mu_B^2+ \kappa_S \hat\mu_S^2 + \kappa_{BS}\hat\mu_B \hat\mu_S  \right),
\label{eq:t}
\end{equation} 
as a finite chemical potential does not alter chiral symmetry breaking. It is thus easy to see that at $\mu_X=0$ the contribution from the singular part of the partition functions to the generalized susceptibilities follows the pattern 
\cite{Ejiri:2005wq}, 
\begin{equation}
\chi^X_{2n}\sim |t|^{2-n-\alpha},
\end{equation}
where $\alpha$ is the critical exponent of the specific heat, which is small and negative. The fourth order cumulants will thus develop a cusp in the chiral limit, whereas the sixth order cumulants are divergent with amplitudes that have different signs below and above $T_c$. The lattice data is qualitatively consistent with that picture as can be seen in Fig.~\ref{fig:chi} (left and middle). A more detailed scaling analysis using a recent parametrization of the $O(4)$-scaling function of the specific heat \cite{O4} is work in progress. It will allow to obtain various non-universal normalization constants that map QCD to the universal $O(4)$-symmetric theory such as $T_0,t_0,h_0,\kappa_X,\kappa_{XY}$. Some of them are of immediate interest, such as the transition temperature in the chiral limit $T_0$ or the curvature $\kappa_X$ that characterizes the change of the transition temperature in the direction of the chemical potential $\mu_X$. Recently the HotQCD Collaboration and the BNL-Bielefeld Collaboration performed a scaling analysis of the chiral condensate, the chiral susceptibility \cite{magnetic, Tc} as well as a mixed susceptibility \cite{kappa}.  That also determined the normalization constant $z_0=h_0^{1/\beta\delta}/t_0$, where $\beta$ and $\delta$ are critical exponents, which fixes the quark mass dependence of the QCD transition (at $\mu_X=0)$. This was used by HotQCD in order to determine the crossover temperature $T_c=154(9)$~MeV \cite{Tc} at physical quark masses.

\subsection{Comparison with the Hadron Resonance Gas \label{sec:HRG}}
Below the QCD transition temperature we can compare our lattice results with the statistical Hadron Resonance Gas (HRG) model, which describes the hadronization process in heavy ion collisions quite successfully \cite{HRG}. For the second order cumulants of baryon number, electric charge and strangeness fluctuations this comparison is performed in Fig.~\ref{fig:chi} (right) \cite{c2}. Here we plot the lattice data normalized by the corresponding HRG results. Light blue bands indicate the continuum extrapolations based on lattices with temporal extent $N_\tau=6,8,12$. We find that the continuum extrapolations of $\chi^B_2$ and $\chi^Q_2$ approach the HRG from below and are in agreement with the HRG at temperatures up to $T\sim (150-160)$ MeV. The strangeness fluctuations, however, seem to overshoot the HRG and eventually approach the HRG at lower temperatures from above. Note that for the electric charge fluctuations additional solid lines at low temperature indicate HRG results with a modified pion mass. We have chosen the pion masses such that they agree with the averaged pion mass (root-mean-square mass $m_\pi^{\rm RMS}$) of the pion spectrum on the lattice at given $N_\tau$. In general, we find that the modified HRG provides a good approximation to the electric charge fluctuations below $T_c$. 

A comparison of the HRG with higher order cumulants is at present not very meaningful as we do not have a continuum result yet.  
Moreover, for the electric charge fluctuations, which can in principle be immediately compared to the experimental results, the distorted pion spectrum on the lattice becomes increasingly problematic, as the higher order cumulants are increasingly sensitive to an increased $m_\pi^{\rm RMS}$. One thus has to use even finer lattice spacings such as $N_\tau=16$ in order to control this systematic effect. In fact, we find that the HRG results for $\chi^Q_4$ with a pion mass corresponding to the $m_\pi^{\rm RMS}$ of our $N_\tau=6$ and $8$ lattices are indistinguishable from that of a HRG with an infinitely heavy pion mass.

\subsection{Determination of freeze-out conditions}
\begin{figure*}[t]
\begin{center}
\includegraphics[width=0.32\textwidth]{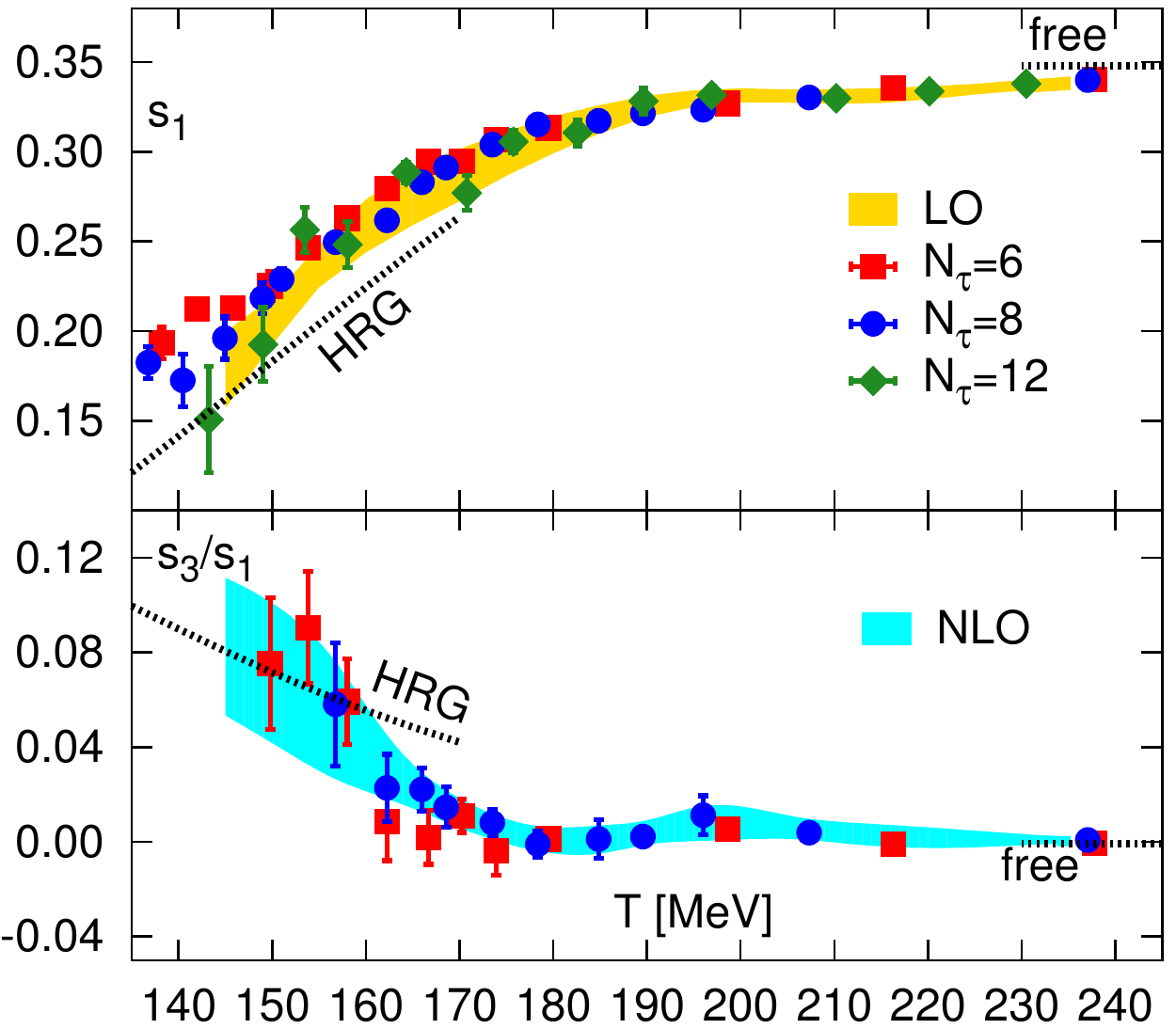} 
\includegraphics[width=0.32\textwidth]{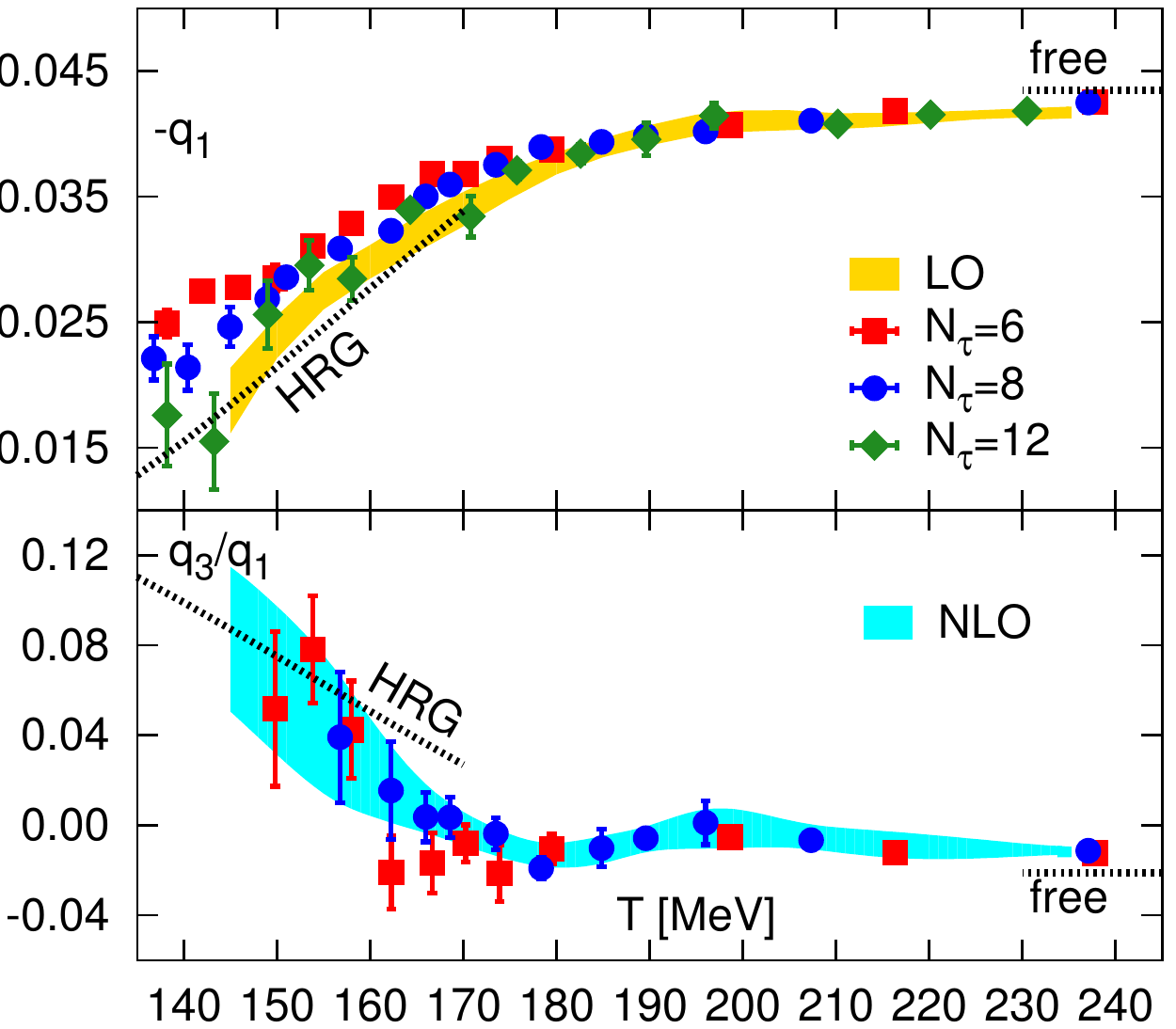} 
\includegraphics[width=0.32\textwidth]{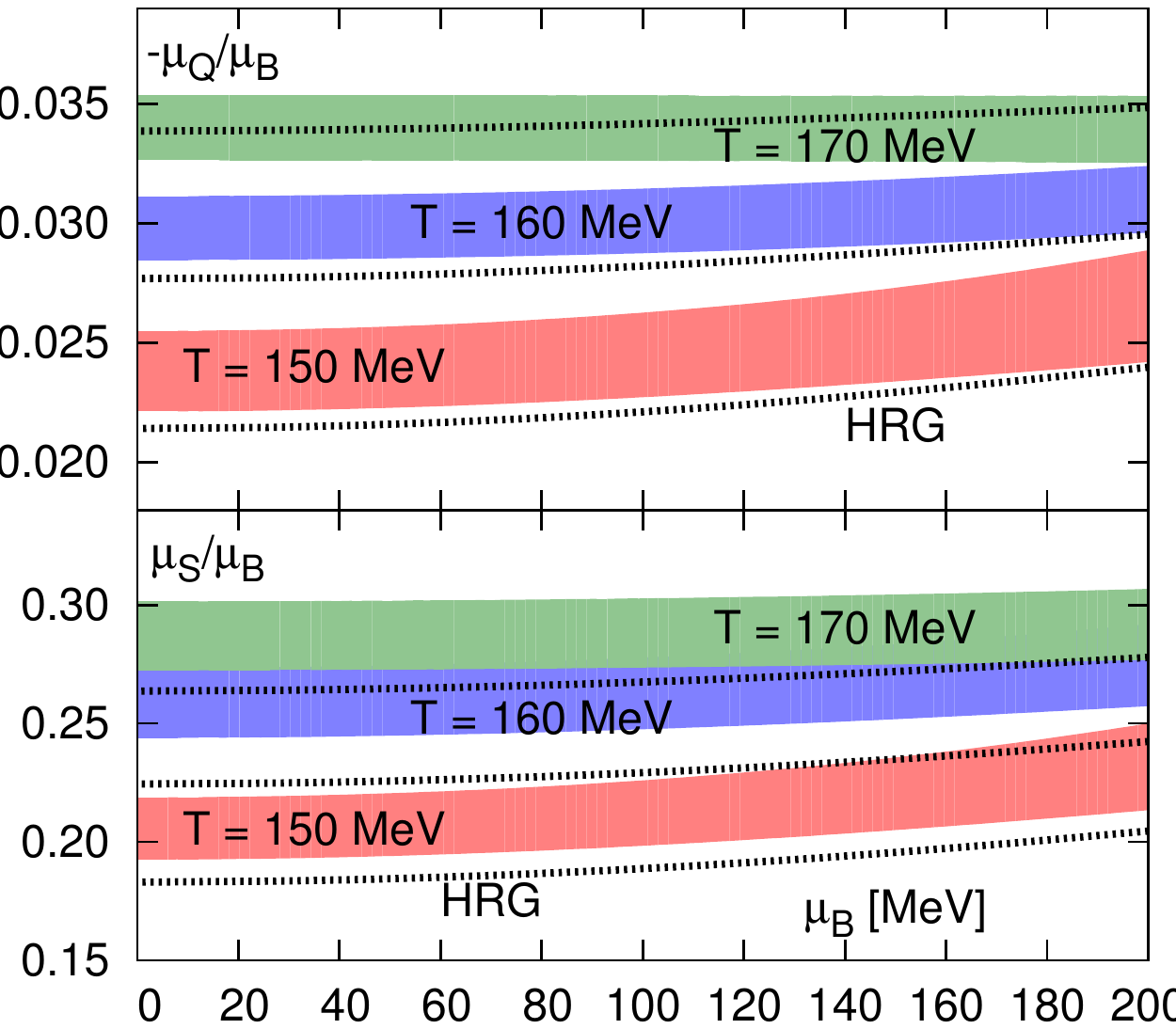} 
\caption{The leading and next-to-leading order expansion coefficients 
of the strangeness (left) and the negative of the electric charge chemical 
potentials (middle) versus temperature for $r=0.4$.
For $s_1$ and $q_1$ the LO-bands show results for the continuum 
extrapolation. For $s_3$ and $q_3$ we give an estimate for continuum 
results (NLO bands) based on spline interpolations of the $N_\tau=8$ data.
Dashed lines at low temperature are from the HRG model
and at high temperature from a massless, 3-flavor quark gas. 
The right hand panel shows NLO results for $\mu_S/\mu_B$ and
$\mu_Q/\mu_B$ as function of $\mu_B$ for three values of the 
temperature. 
\vspace*{-0.5cm}
}
\label{fig:chempot}
\end{center}
\end{figure*}

In oder to compare the lattice QCD results with experimentally measured fluctuations one has to eliminate the unknown fireball volume by taking ratios of cumulants. We have recently proposed a detailed method to extract the remaining freeze-out parameters, {\it i.e.}  the freeze-out temperature ($T^f$) and the freeze-out chemical potential ($\mu_B^f$), from a comparison of lattice and experimental results of ratios of cumulants of the electric charge fluctuations \cite{PRL,Swagato}. In order to resemble the conditions met in heavy ion conditions we fix the strangeness and electric charge chemical potentials by demanding net strangeness neutrality and the correct iso-spin asymmetry:
\begin{equation} 
\left< N_S\right>=0,\qquad \left< N_Q\right>=r\left< N_B\right>.
\label{eq:cond}
\end{equation}
Here $r$ is a constant. We choose $r = 0.4$, which approximates well the situation met in Au-Au and Pb-Pb collisions. We fulfill the conditions given in Eq.~\ref{eq:cond} by expanding the net densities as well as the chemical potentials $\hat \mu_S$ and $\hat \mu_Q$ in terms of $\hat \mu_B$. We then solve order by order for the expansion coefficients of the chemical potentials. In next-to-leading order (NLO) we obtain
\begin{equation}
\hat\mu_S=s_1(T)\hat\mu_B+s_3(T)\hat\mu_B^3+\mathcal{O}(\hat\mu_B^5),\qquad
\hat\mu_Q=q_1(T)\hat\mu_B+q_3(T)\hat\mu_B^3+\mathcal{O}(\hat\mu_B^5),
\end{equation}
where the results for $s_1(T)$, $s_3(T)$, $q_1(T)$, $q_3(T)$ can easily be expressed in terms of the generalized susceptibilities $\chi^{BQS}_{ijk}$. In Fig.~\ref{fig:chempot} (left and middle) we show those expansion coefficients. 
The upper panel shows the leading order (LO), whereas the lower panels show the ratio NLO to LO coefficients. The band in the upper panels indicates the continuum extrapolation based on the $N_\tau=6,8$ and 12 data, the band in the lower panels is a continuum estimate based on $N_\tau=6$ and 8 data. We find that the NLO contributions are negligible in the high temperature region and below 10\% in the temperature interval relevant for the analysis of freeze-out conditions, {\it i.e.}, $T\approx (160 \pm 10)$ MeV. In fact, in this temperature range the leading order lattice QCD results deviate from HRG model calculations expanded to the same order by less than 15\%. Note that one can also investigate the convergence properties of the HRG model itself.  In the HRG model the NLO expansion reproduces the full HRG result for $\hat\mu_Q$ and $\hat\mu_S$ to better than 1.0\% for all values of $\hat\mu_B\lesssim 1.3$. Altogether, we thus expect that the NLO truncated QCD expansion is a good approximation to the complete QCD results for $\hat\mu_Q$ and $\hat\mu_S$ for $\mu_B \lesssim 200$ MeV.

Our results for the strangeness and electric charge chemical potentials at NLO as function of  $\mu_B$ and $T$ are shown in Fig.~\ref{fig:chempot} (right). While $\mu_S/\mu_B$ varies between 0.2 and 0.3 in the interval 150 MeV $\lesssim T \lesssim 170$ MeV, the absolute value of $\mu_Q/\mu_B$ is an order of magnitude smaller. Both ratios are almost constant for $\mu_B \lesssim 200$ MeV, which is consistent with HRG model calculations.

We do now consider ratios of cumulants $R^X_{nm} = \chi^X_{n} /\chi^X_{m} $, which to a large extent eliminate the dependence of cumulants on the freeze-out volume. After fixing $\mu_Q$ and $\mu_S$ by the constraints given in Eq.~\ref{eq:cond}, they only depend on $\mu_B$ and $T$. Ratios with $n + m$ even are non-zero for $\mu_B = 0$, while the odd-even ratios are in leading order proportional $\mu_B$ and thus vanish for $\mu_B=0$. Ratios with $n+m$ even or odd thus provide complementary information on $T^f$ and $\mu^f_B$. The simplest of those ratios are given by 
\begin{eqnarray}
\hspace*{-0.5cm}R_{12}^X &\equiv& 
\frac{M_X}{\sigma_X^2} = 
\hat\mu_B \left(
R_{12}^{X,1} + R_{12}^{X,3}\ \hat\mu_B^2 + {\cal O}(\hat\mu_B^4)
\right)\; ,
\label{R12} \\
\hspace*{-0.5cm}R_{31}^X &\equiv& 
\frac{S_X \sigma_X^3}{M_X} = 
 R_{31}^{X,0} + R_{31}^{X,2}\ \hat\mu_B^2 + {\cal O}(\hat\mu_B^4)\; .
\label{R31}
\end{eqnarray}
These ratios can be calculated in QCD as well as in the HRG model \cite{Ejiri:2005wq}, and eventually can be compared to experimental data in order to determine $T^f$ and $\mu_B^f$ . Here we have evaluated them up to the NLO contribution $\mathcal{O}(\hat\mu_B^3)$ in the Taylor expansion of $R_{12}^X$  and to LO for $R_{31}^X$. For the discussion of results presented below, we concentrate on net electric charge fluctuations, {\it i.e.} on $R_{12}^Q$ and $R_{31}^Q$. While net baryon number fluctuations are experimentally accessible only through measurements of net proton number fluctuations, which may cause some difficulties \cite{Kitazawa,Bzdak}, electric charge fluctuations may be easier to analyze. They are thus better suited for a comparison with the experimental data. 

\begin{figure*}[t]
\begin{center}
\includegraphics[width=0.32\textwidth]{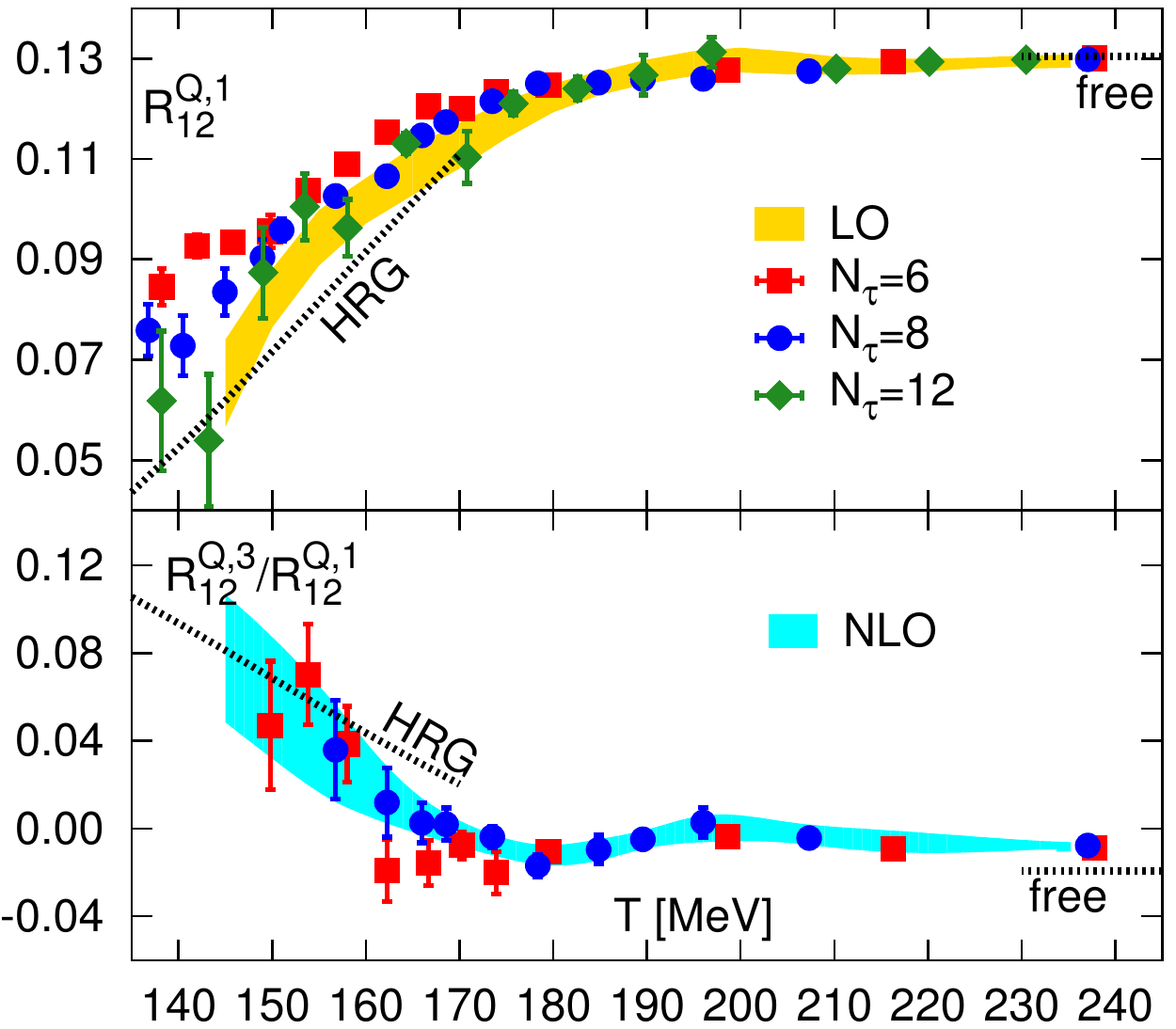}
\includegraphics[width=0.32\textwidth]{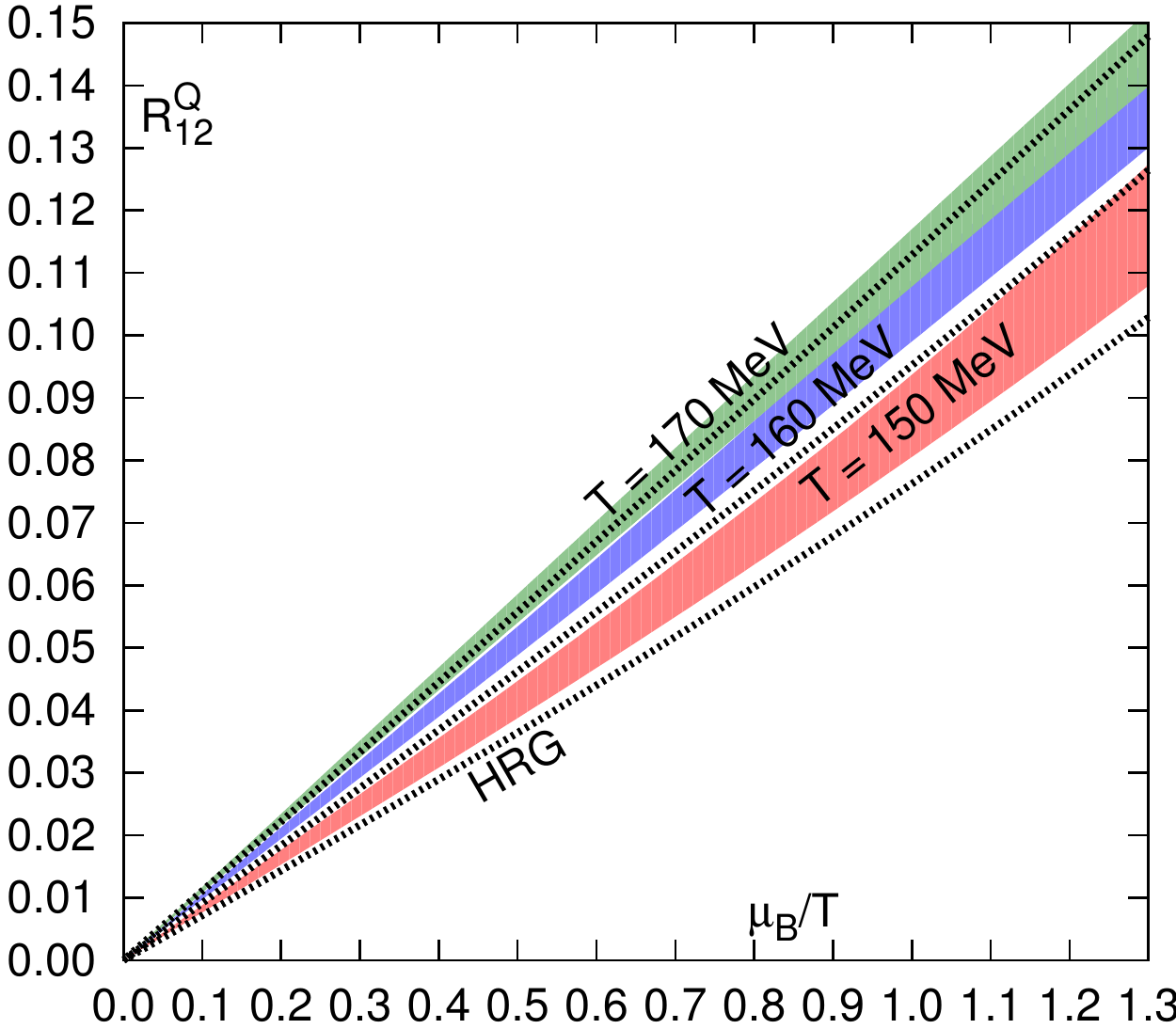}
\includegraphics[width=0.32\textwidth]{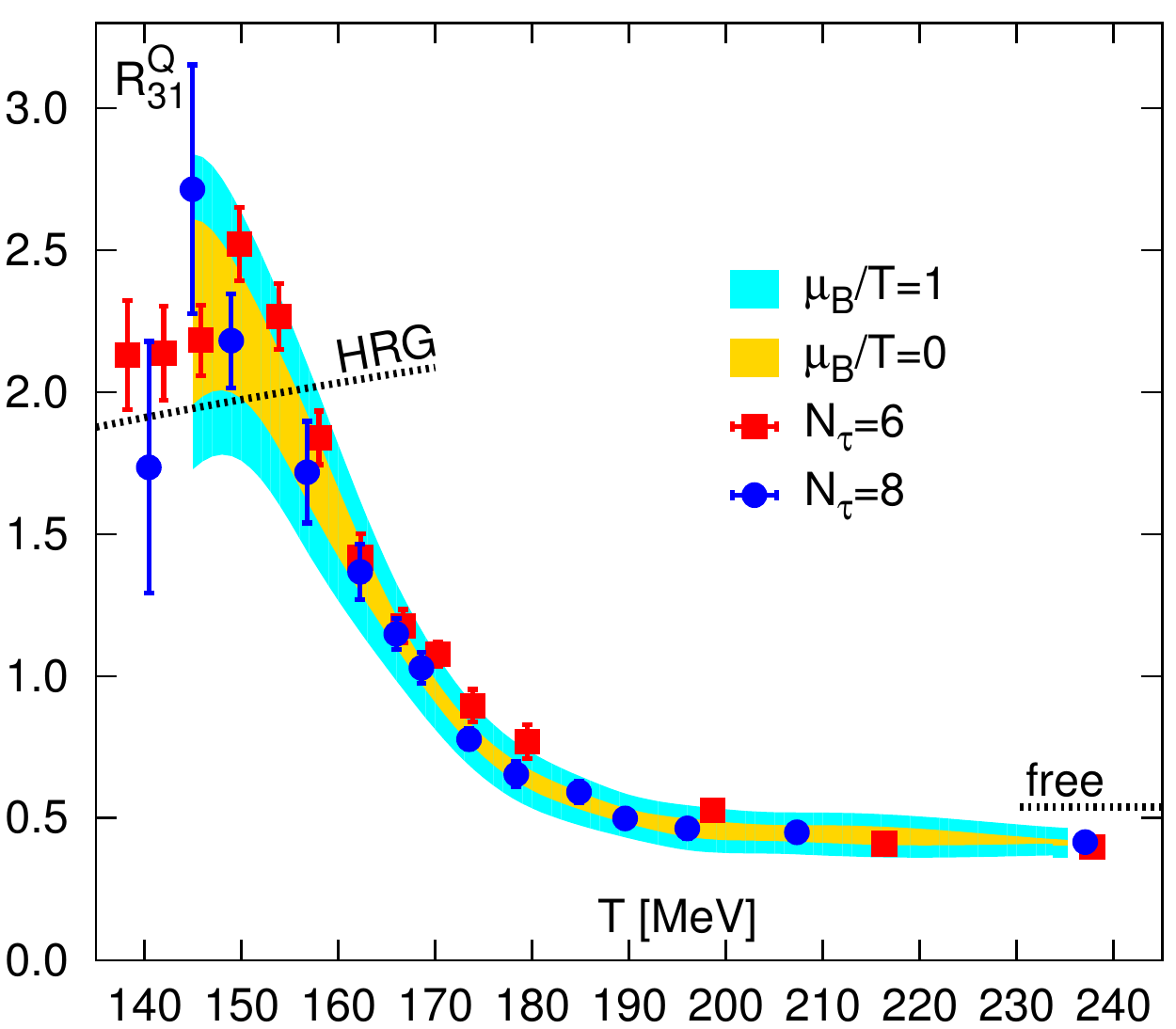}
\caption{The left panel shows LO (top) and NLO (bottom) expansion coefficients of $R_{12}^Q$ for for $r=0.4$. The bands and lines are as in 
Fig.~\protect\ref{fig:chempot}(left).On the middle panel we plot $R_{12}^Q$ versus $\mu_B/T$, including the NLO contribution, for three
values of the temperature. $R_{31}^Q$ versus temperature is shown on the right panel for $\mu_B=0$. The wider band on the data set for 
$N_\tau=8$ shows an estimate of the magnitude of NLO corrections. }
\label{fig:R31}
\end{center}
\vspace*{-0.5cm}
\end{figure*}

In Fig.~\ref{fig:R31} (left) we show our results on the LO and NLO expansion coefficients of $R_{12}^Q$. The bands on the lower and upper panel have the same meaning as in Fig.~\ref{fig:chempot} (left). 
We find that the NLO corrections to $R_{12}^Q$ are below 10\%, which makes the LO result a good approximation for a large range of $\hat\mu_B$. Systematic errors arising from the truncation of the Taylor series for $R_{12}^Q$ at NLO may again be estimated by comparing the full result in the HRG model calculation with the corresponding truncated results. Here we find for $T = (160 \pm 10)$ MeV and $\hat\mu_B\lesssim 1.3$ that the difference is less than 1.0\%. Moreover, we estimated that taste violation effects in the NLO calculation lead to systematic errors that are at most 5\% and thus will be negligible in $R_{12}^Q$ . Taylor series truncated at NLO are thus expected to give a good approximation to the full result for a wide range of baryon chemical potentials.

In Fig.~\ref{fig:R31} (middle) we show the full $\mu_B$ and $T$ dependence of $R_{12}^Q$, including the NLO contribution. Obviously the ratio $R_{12}^Q$ shows a strong sensitivity on $\mu_B$ but varies little with $T$ in the temperature range $T=(160\pm 10)$ MeV. For the determination of $(T^f,\mu_B^f)$ a second, complimentary information is needed. To this end we use the ratio $R_{31}^Q$, which is strongly dependent on $T$ but receives corrections only at $\mathcal{O}(\hat\mu_B^2)$. The leading order result for this ratio is shown in Fig.~\ref{fig:R31} (right).  Apparently this ratio shows a characteristic temperature dependence for $T\gtrsim 155$ MeV that is quite different from that of HRG model calculations. The NLO correction to this ratio vanishes in the high temperature limit and at low $T$ the HRG model also suggests small corrections. In fact, in the HRG model the LO contributions to $R_{31}^Q $ differ by less than 2\% from the exact results on the freeze-out curve for $\mu_B \lesssim 200$ MeV. The broader band in Fig.~\ref{fig:R31} (right) indicates an estimate of the NLO contribution at $\hat\mu_B=1$ from our $N_\tau=8$ calculations.

We now are in the position to extract $\mu_B^f$ and $T_f$ from $R_{12}^Q$ and $R_{31}^Q$ which eventually will be measured in the beam energy scan at RHIC \cite{Kumar}. A large value for $R_{31}^Q$, {\it i.e.} $R_{31}^Q \simeq 2$ would suggest a low freeze-out temperature $T \lesssim 155$ MeV, while a value $R_{31}^Q \simeq 1$ would suggest a large freeze-out temperature, $T \sim 170$ MeV. A value of $R_{31}^Q \simeq 1.5$ would correspond to $T\sim 160$ MeV.
A measurement of $R_{31}^Q$ thus suffices to determine the freeze-out temperature. In the HRG model parametrization of the freeze-out curve \cite{fc} the favorite value for $T^f$ in the beam energy range 200 GeV $\geq \sqrt{s_{NN}} \geq 39$ GeV indeed varies by less than 2 MeV and is about 165 MeV.
At this temperature the values for $R_{31}^Q$ calculated in the HRG model and in QCD differ quite a bit, as is obvious from Fig.~\ref{fig:R31} (right). While $R_{31}^Q \simeq 2$ in the HRG model, one finds $R_{31}^Q \simeq 1.2$ in QCD at $T=165$ MeV. Values close to the HRG value are compatible with QCD calculations only for $T\lesssim 157$ MeV. We thus expect to either find freeze-out temperatures that are about 5\% below HRG model results or values for $R_{31}^Q$ that are significantly smaller than the HRG value. A measurement of this cumulant ratio at RHIC thus will allow to determine $T^f$ and probe the consistency with HRG model predictions.

For any of these temperature values a comparison of an experimental value for $R_{12}^Q$ with Fig.\ref{fig:R31} (middle)  will allow to determine $\mu_B^f$. To be specific, at $T = 160$ MeV we expect to find $\mu_B^f = (20 - 30)$ MeV, if $R_{12}^Q$ lies in the range $0.012 - 0.020$, $\mu_B^f = (50 - 70)$ MeV for $0.032 \leq R_{12}^Q \leq 0.045$ and $\mu_B^f = (80 -  120)$ MeV for $0.05 \leq R_{12}^Q \leq 0.08$. These parameter ranges are expected \cite{fc} to cover the regions relevant for RHIC beam energies $\sqrt{s_{NN}} = 200$ GeV, 62.4 GeV and 39 GeV, respectively. As is evident from Fig.~\ref{fig:R31} (middle) the values for $\mu_B^f$ will shift to smaller (larger) values when $T^f$ turns out to be larger (smaller) than 160 MeV. A more refined analysis of $(T^f , \mu_B^f)$ will become possible, once the ratios $R_{12}^Q$ and $R_{31}^Q$ have been measured experimentally.

\section{Summary}
We have reviewed the current status of our equation of state calculations with the HISQ action, based on lattices of temporal extension $N_\tau=6,8,10,12$.  Recent results for the trace anomaly indicate a discrepancy of roughly 20\% in the peak region ($T\approx 200$ MeV) between HotQCD \cite{Bazavov} and the Budapest-Wuppertal results \cite{stout}. However, final conclusions can only be drawn after the continuum extrapolation of our data, which we will attempt as soon as the $N_\tau=10$ and 12 data have reached higher statistical significance. We have further discussed fluctuations of conserved charges. We have shown that higher order cumulants of conserved charge fluctuations become more and more sensitive to critical behavior. For the second order fluctuations we have explicitly discussed their approach to the HRG in the low temperature region. Finally, we proposed a method to extract freeze-out parameters, {\it i.e.}  the freeze-out temperature ($T^f$) and the freeze-out chemical potential ($\mu_B^f$), from a comparison of lattice and experimental results of ratios of cumulants of the electric charge fluctuations \cite{PRL,Swagato}. At nonzero baryon density the method is based on a next to leading order (NLO) Taylor expansion of two different ratios. In general we find that the NLO contributions are below 10\% in the $\mu_B/T$ range relevant for BES at RHIC down to collision energies of $\sqrt{s_{NN}}\gtrsim 20$~GeV.

\section*{Acknowledgments}
The numerical generation of gauge field configurations has been performed on BlueGene/L computers at Lawrence Livermore National Laboratory (LLNL), the New York Center for Computational Sciences (NY-CCS) at Brookhaven National Laboratory, US Teragrid (Texas Advanced Computing Center), Cray XE6 at the National Energy Research Scientific Computing Center (NERSC), and on clusters of the USQCD collaboration in JLab and FNAL. The evaluations of fluctuation observables have been performed on the USQCD owned GPU-cluster at JLab and the GPU-cluster at the University of Bielefeld. We further acknowledge support by contract DE-AC02-98CH10886 with the U.S. Department of Energy, the Bundesministerium f\"ur Bildung und Forschung under grant 06BI9001, the Gesellschaft f\"ur Schwerionenforschung under grant BILAER, the Deutsche Forschungsgemeinschaft under grant GRK881 and the EU Integrated Infrastructure Initiative Hadron-Physics 2.


\section*{References}
\begin{thebibliography}{99}
\bibitem{Romatschke} P.~Romatschke and U.~Romatschke, Phys.\ Rev.\ Lett.\  {\bf 99}, 172301 (2007) [arXiv:0706.1522 [nucl-th]]. 
\bibitem{Schenke} C.~Gale, S.~Jeon, B.~Schenke, P.~Tribedy and R.~Venugopalan,  arXiv:1210.5144 [hep-ph].
\bibitem{swansea} C.~R.~Allton, {\it et al.}, Phys.\ Rev.\ D {\bf 66} (2002) 074507 [hep-lat/0204010].
\bibitem{epj} C.~Schmidt, Eur.\ Phys.\ J.\ C {\bf 61}, 537 (2009).
\bibitem{rajagopal} M.~A.~Stephanov, K.~Rajagopal and E.~V.~Shuryak, Phys.\ Rev.\ D {\bf 60}, 114028 (1999) [hep-ph/9903292].
\bibitem{PRL} A.~Bazavov {\it et al.}, arXiv:1208.1220 [hep-lat], to appear in Phys.\ Rev.\ Lett.
\bibitem{Swagato} S.~Mukherjee, Quark Matter 2012.
\bibitem{HRG} P. Braun-Munzinger, K. Redlich, J. Stachel, In *Hwa, R.C. (ed.) et al.: Quark gluon plasma*, 491-599.
\bibitem{fc} J.~Cleymans, H.~Oeschler, K.~Redlich and S.~Wheaton, Phys.\ Rev.\ C {\bf 73}, 034905 (2006).
\bibitem{HISQ} E.~Follana {\it et al.}  [HPQCD and UKQCD Collaborations], Phys.\ Rev.\ D {\bf 75}, 054502 (2007).
\bibitem{Sommer} R.~Sommer, Nucl.\ Phys.\ B {\bf 411}, 839 (1994) [hep-lat/9310022].
\bibitem{Bazavov}  A.~Bazavov [HotQCD Collaboration], arXiv:1210.6312 [hep-lat].
\bibitem{stout} S. Borsanyi {\it et al.}, [BW Collaboration], JHEP {\bf 1011}, 077 (2010), [arXiv:1007.2580[hep-lat]].
\bibitem{Ejiri:2005wq} S.~Ejiri, F.~Karsch, K.~Redlich, Phys.\ Lett.\  {\bf B633}, 275 (2006).
\bibitem{O4} J.~Engels and F.~Karsch, Phys.\ Rev.\ D {\bf 85}, 094506 (2012) [arXiv:1105.0584 [hep-lat]].
\bibitem{magnetic} S. Ejiri {\it et al.}, Phys.\ Rev.\ D {\bf 80}, 094505 (2009) [arXiv:0909.5122 [hep-lat]].
\bibitem{Tc} A.~Bazavov {\it et al.}, Phys.\ Rev.\ D {\bf 85}, 054503 (2012) [arXiv:1111.1710 [hep-lat]].
\bibitem{kappa} O.~Kaczmarek {\it et al.}, Phys.\ Rev.\ D {\bf 83}, 014504 (2011) [arXiv:1011.3130 [hep-lat]].
\bibitem{c2} A.~Bazavov, {\it et al.}, Phys.\ Rev.\ D {\bf 86}, 034509 (2012) [arXiv:1203.0784 [hep-lat]].
\bibitem{STAR_B} M.~M.~Aggarwal {\it et al.} [STAR Collaboration], Phys.\ Rev.\ Lett.\ 105, 022302 (2010).
\bibitem{Kitazawa} M.~Kitazawa, M.~Asakawa, Phys.\ Rev.\ C {\bf 85}, 021901 (2012).
\bibitem{Bzdak} A.~Bzdak, V.~Koch, V.~Skokov, arXiv:1203.4529 [hep-ph].
\bibitem{Kumar}L.~Kumar [STAR Collaboration], Quark Matter 2012.
\end{thebibliography}
\end{document}